\documentclass[conference]{IEEEtran}
\PassOptionsToPackage{ruled, linesnumbered}{algorithm2e}
\usepackage[T1]{fontenc}
\usepackage[utf8]{inputenc}
\usepackage{cprotect}
\usepackage{units}
\usepackage{algorithm2e}
\usepackage{amsmath}
\usepackage{amssymb}
\usepackage{graphicx}
\usepackage[bookmarks=true,bookmarksnumbered=true,bookmarksopen=true,bookmarksopenlevel=1,
 breaklinks=false,pdfborder={0 0 0},pdfborderstyle={},backref=false,colorlinks=false]
 {hyperref}
\hypersetup{pdftitle={Your Title},
 pdfauthor={Your Name},
 pdfpagelayout=OneColumn, pdfnewwindow=true, pdfstartview=XYZ, plainpages=false}

\makeatletter

\usepackage[caption=false,font=footnotesize]{subfig}
\usepackage{amsmath, amssymb, amsfonts, graphicx}

\allowdisplaybreaks
\newcommand{\herm}{^{\mathsf{H}}}
\newcommand{\trans}{^{\mathsf{T}}}

\DeclareMathOperator{\maximize}{maximize}
\DeclareMathOperator{\st}{subject~to}

\makeatother

\begin{document}
\title{{\huge Stacked Flexible Intelligent Metasurface Design for Multi-User Wireless Communications} \vspace{-0.4cm}}
\author{\IEEEauthorblockN{Ahmed~Magbool\IEEEauthorrefmark{1}, Vaibhav~Kumar\IEEEauthorrefmark{2}, Marco~Di~Renzo\IEEEauthorrefmark{3}\IEEEauthorrefmark{4}, and Mark~F.~Flanagan\IEEEauthorrefmark{1}}\IEEEauthorblockA{\IEEEauthorrefmark{1}School of Electrical and Electronic Engineering, University College Dublin, Belfield, Dublin 4, Ireland\\
\IEEEauthorrefmark{2}Wireless Research Lab, Engineering Division, New York University Abu Dhabi (NYUAD), Abu Dhabi, UAE\\
\IEEEauthorrefmark{3}Université Paris-Saclay, CNRS, CentraleSupélec, Laboratoire des Signaux et Systémes, Gif-sur-Yvette, France\\
\IEEEauthorrefmark{4}Centre for Telecommunications Research, Department of Engineering, King's College London, UK\\
Email: ahmed.magbool@ucdconnect.ie, vaibhav.kumar@ieee.org, marco.di\_renzo@kcl.ac.uk, mark.flanagan@ieee.org }\vspace{-0.9cm} }
\maketitle
\begin{abstract}
Stacked intelligent metasurfaces (SIMs) have recently emerged as an effective solution for next-generation wireless networks. A SIM comprises multiple metasurface layers that enable signal processing directly in the wave domain. Moreover, recent advances in \emph{flexible metamaterials} have highlighted the potential of flexible intelligent metasurfaces (FIMs), which can be physically morphed to enhance communication performance. In this paper, we propose a \emph{stacked flexible intelligent metasurface} (SFIM)-based communication system for the first time, where each metasurface layer is deformable to improve the system’s performance. We first present the system model, including the transmit and receive signal models as well as the channel model, and then formulate an optimization problem to maximize the system sum rate under constraints on the transmit power budget, morphing distance, and the unit-modulus condition of the meta-atom responses. To solve this problem, we develop an alternating optimization framework based on the gradient projection method. Simulation results demonstrate that the proposed SFIM-based system achieves significant performance gains compared to its rigid SIM counterpart.
\end{abstract}


\IEEEpeerreviewmaketitle{}

\section{Introduction\label{sec:Introduction}}

Massive multiple-input multiple-output (mMIMO) is an enabling technology for sixth-generation (6G) networks, offering sharp beamforming capabilities and enhanced spatial diversity. However, it relies on large-scale phased arrays with high-resolution responses, leading to increased cost and power consumption~\cite{2014_Larsson}. 

To address these limitations, recent studies have introduced the concept of stacked intelligent metasurfaces (SIMs), consisting of multiple metasurface layers that enable signal processing directly in the wave domain~\cite{2025_Liu}. In contrast to conventional mMIMO transceivers, SIM-based architectures provide a cost-efficient alternative by replacing power-hungry digital baseband modules with a metasurface structure. The utilization of SIM for multi-user communication was first introduced in~\cite{2023_An}, where the authors demonstrated that SIM-based systems can significantly reduce the precoding delay of conventional digital beamforming architectures. The study in~\cite{2023_An1} further showed that SIM-based systems can achieve up to a 150\% increase in capacity compared to both conventional MIMO and reconfigurable intelligent surface (RIS)-assisted systems. The authors in~\cite{2025_Papazafeiropoulos} proposed a system model with two SIMs: one integrated into the transmitter and the other placed in front of the users, while considering statistical channel state information (CSI) to alleviate the burden of re-optimizing the system within each coherence time. Similarly, a two-SIM system was investigated in~\cite{2024_Papazafeiropoulos}, where the achievable rate was maximized for a system in which both the transmitter and receiver were equipped with SIMs. Several other works have explored different SIM system models and scenarios, such as physical-layer security~\cite{2024_Niu,2025_Kavianinia}, cell-free multiple-input multiple-output (MIMO) networks~\cite{2024_Li,2025_Hu}, wireless sensing~\cite{2025_Teng}, and meta-atoms exhibiting amplification properties~\cite{2025_Darsena}.

Moreover, a flexible intelligent metasurface (FIM) is a reconfigurable and deformable metasurface whose electromagnetic properties and physical geometry can be jointly controlled to dynamically manipulate wireless propagation~\cite{2025_An3}. Such surfaces offer significant advantages over conventional rigid surfaces, enabling them to maintain optimal propagation characteristics under varying environmental conditions. The work in~\cite{2025_An} demonstrated that a base station (BS) equipped with a transmit FIM can reduce power consumption by half compared compared to rigid arrays under the same quality-of-service (QoS) requirements. The authors of~\cite{2025_An1} investigated the maximization of the channel capacity in a point-to-point MIMO system employing FIMs at both the transmitter and receiver. They showed that the channel capacity can be doubled in a FIM-based system compared to a system with a rigid array. The use of a FIM as a reflector was studied in~\cite{2025_Hu3}, where the channel gain was maximized. This study showed that the flexibility of FIMs can significantly enhance the channel gain compared to conventional rigid RISs.

In this paper, for the first time, we investigate a system that combines both of these concepts where we employ a SIM at the transmitter side with morphable layers, referred to as a \textit{stacked FIM} (SFIM). Our objective is to jointly design the power allocation, surface morphing, and meta-atom responses to maximize the system sum rate, subject to constraints on the transmit power budget, morphing range of the meta-atoms, and unit-modulus condition of the meta-atom responses. To tackle the resulting challenging optimization problem, we employ an alternating optimization (AO) framework combined with a gradient projection method. Our simulations show that the proposed SFIM architecture provides a significant sum rate gain compared to conventional rigid SIM (RSIM)-based systems.

\cprotect\section{System Model \label{sec:sys_model} 
\begin{figure}[tb]
\protect\centering{}\protect\includegraphics[width=0.75\columnwidth]{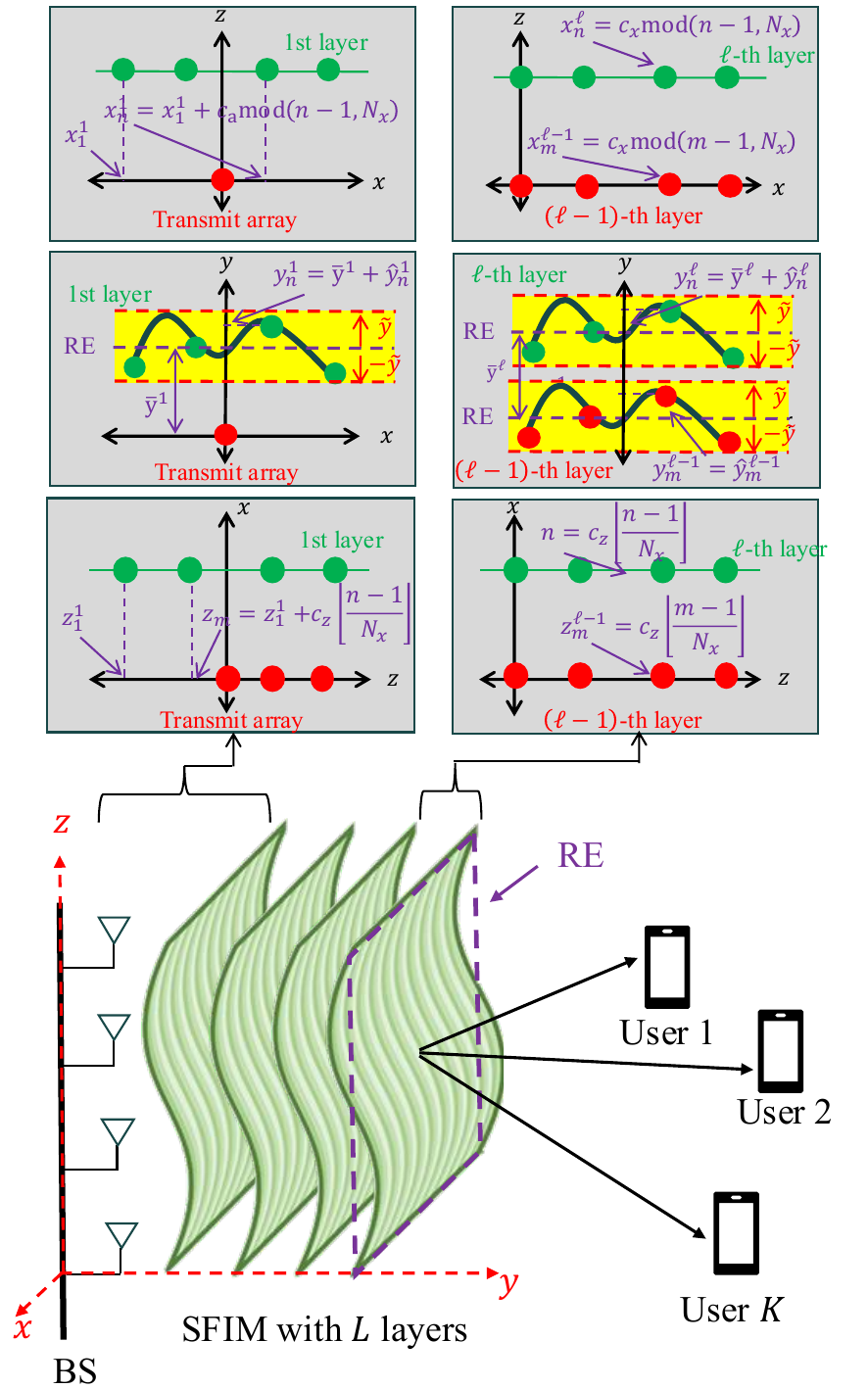} \protect\caption{Proposed system model with SFIM comprising $L$ layers serving $K$ users. The figure also illustrates the geometry of the transmitter-to-first-layer distances and the inter-layer distances.}
\label{fig:sys_model}\protect
\end{figure}
}

We consider downlink communication between a BS, equipped with $M$ antennas and an SFIM, and $K$ single-antenna users, as depicted in Fig.~\ref{fig:sys_model}. The SFIM comprises $L$ layers of FIMs, each comprising $N$ meta-atoms. In the following, we present the transmit signal model, channel model and received signal model.

\subsection{Transmit Signal Model}

The BS transmits the signal $\mathbf{x}=\mathbf{P}\mathbf{s},$ where $\mathbf{P}\triangleq\text{diag}(\sqrt{p_{1}},\dots,\sqrt{p_{K}})$, with $p_{k}$ representing the power allocated to the $k$-th user, and $\mathbf{s}\triangleq[s_{1},\dots,s_{k}]\trans$ with $s_{k}\in\mathbb{C}$ denoting the $k$-th user's symbol with $\mathbb{E}\{\mathbf{s}\mathbf{s}\herm\}\sim\mathcal{CN}(\mathbf{0},\mathbf{I})$.

\subsection{Channel Model}

Without loss of generality, we assume that the transmit array is located along the $z$-axis as shown in Fig.~\ref{fig:sys_model}. We then define the \textit{rigid equivalent} (RE) of the $\ell$-th metasurface as a reference surface that is parallel to the $x$–$z$ plane, where all meta-atoms share the same $y$-coordinate\footnote{This convention is used for simplicity. However, other orientations of the transmit antennas and the REs of SFIM layers can also be considered.}. Furthermore, we define the distance along the $y$-axis between the transmit antennas and the RE of the first layer as $\bar{y}^{1}$. Similarly, we define the distance along the $y$-axis between the REs of the $(\ell-1)$-th layer and the $\ell$-th layer as $\bar{y}^{\ell}$ for all $\ell\in\mathcal{L}\setminus\{1\}$ with $\mathcal{L}\triangleq\{1,\dots,L\}$. We also define $[-\tilde{y},\tilde{y}]$ as the \textit{morphing range} for each meta-atom, representing the range within which the meta-atom can move along the $y$-axis\footnote{The maximum morphing distance should ensure that elements on two different surfaces are not too close to each other. In general, the distance between elements in different layers should account for the fact that transmission should not occur in the reactive near field, so that the Rayleigh–Sommerfeld channel model remains valid.}. This range is assumed to be symmetric around $\bar{y}^{\ell}$ for all $\ell \in \mathcal{L}$.

The cascaded channel between the transmit antenna array at the BS and the $k$-th user can be expressed as 
\begin{equation}
\mathbf{g}_{k}\trans(\hat{\mathbf{y}},\boldsymbol{\phi})=\mathbf{h}_{k}\trans(\hat{\mathbf{y}})\boldsymbol{\Phi}^{L}\boldsymbol{\Omega}^{L}(\hat{\mathbf{y}})\dots\boldsymbol{\Phi}^{1}\boldsymbol{\Omega}^{1}(\hat{\mathbf{y}}),\label{eq:ccc}
\end{equation}
where $\boldsymbol{\Phi}^{\ell}\triangleq\text{diag}(\phi_{1}^{\ell},\dots,\phi_{N}^{\ell})$ represents the response of the $\ell$-th SFIM layer with $|\phi_{n}^{\ell}|=1$ for all $\ell\in\mathcal{L}$ and $n\in\mathcal{N}$, $\mathcal{N}\triangleq\{1,\dots,N\}$ and $\boldsymbol{\phi}\triangleq[\phi_{1}^{1},\dots,\phi_{N}^{1},\dots,\phi_{1}^{L},\dots,\phi_{N}^{L}]\trans$. Also, $\mathbf{h}_{k}\trans(\hat{\mathbf{y}})$ is the channel between the final SFIM layer and the $k$-th user, $\boldsymbol{\Omega}^{\ell}(\hat{\mathbf{y}})$ is the channel between the $(\ell-1)$-th layer and the $\ell$-th layer of the SFIM for all $\ell\in\mathcal{L}\setminus\{1\}$, $\boldsymbol{\Omega}^{1}  (\hat{\mathbf{y}})$ is the channel between the transmit antenna array and the first SFIM layer, $\hat{y}_{n}^{\ell}$ is the morphing distance of the $n$-th element in the $\ell$-th layer and $\hat{\mathbf{y}}\triangleq[\hat{y}_{1}^{1},\dots,\hat{y}_{N}^{1},\dots,\hat{y}_{1}^{L},\dots,\hat{y}_{N}^{L}]\trans$.

Using the Rayleigh-Sommerfeld diffraction theory, the channel between the $m$-th transmitting meta-atom in the $(\ell-1)$-th layer (or the $m$-th transmit antenna at the BS when $\ell=1$) and the $n$-th receiving meta-element in the $\ell$-th SFIM layer is expressed as 
\begin{equation}
[\boldsymbol{\Omega}^{\ell}(\hat{\mathbf{y}})]_{n,m}=\tfrac{\tilde{A}^{\ell}\cos(\theta_{n,m}^{\ell})}{d_{n,m}^{\ell}(\hat{\mathbf{y}})}\Big(\tfrac{1}{2\pi d_{n,m}^{\ell}(\hat{\mathbf{y}})}-\tfrac{j}{\lambda}\Big)\exp\Big(j\tfrac{2\pi d_{n,m}^{\ell}(\hat{\mathbf{y}})}{\lambda}\Big),\label{eq:cm1}
\end{equation}
where 
\begin{equation}
\tilde{A}^{\ell}\triangleq\begin{cases}
A_{\text{a}}, & \ell=1,\\
A_{\text{m}}, & \ell\in \mathcal{L} \setminus \{1\},
\end{cases}
\end{equation}
where $A_{\text{a}}$ and $A_{\text{m}}$ represent the surface areas of a transmit antenna and a meta-atom, respectively, $\lambda$ denotes the carrier wavelength, $d_{n,m}^{\ell}(\hat{\mathbf{y}})$ is the distance between the $m$-th transmit meta-atom (or transmit antenna when $\ell=1$) in the $(\ell-1)$-th layer and the $n$-th receiving meta-element in the $\ell$-th layer, and $\theta_{n,m}^{\ell}$ is the angle between the normal direction and the propagation direction between the $m$-th transmit meta-atom (or transmit antenna when $\ell=1$) in the RE of the $(\ell-1)$-th layer and the $n$-th receiving meta-element in the RE of the $\ell$-th layer. Next, we express the channels in terms of the morphing distance for each meta-atom.

\paragraph{\textit{$\boldsymbol{\Omega}^{1} (\hat{\mathbf{y}})$}}

Taking the first antenna at the BS as a reference point, the coordinates of the $m$-th transmit antenna are $\big[x_{m}^{0},\ y_{m}^{0},\ z_{m}^{0}\big]=\big[0,\ 0,\ (m-1)c_{\mathrm{a}}\big]$, where $c_{\mathrm{a}}$ is the spacing between two antenna elements.

Moreover, the coordinate of the $n$-th element of the first flexible metasurface is $\big[x_{n}^{1},\ y_{n}^{1},\ z_{n}^{1}\big]=\big[x_{1}^{1}+c_{x}\mod (n-1,N_{x}),\ \bar{y}^{1}+\hat{y}_{n}^{1},\ z_{1}^{1}+c_{z}\lfloor(n-1)/N_{x}\rfloor\big]$. Here, $x_{1}^{1}$ and $z_{1}^{1}$ represent the $x$- and $z$-coordinates of the reference SFIM meta-atom in the RE, which is assumed to be the element located at the top-left corner, with respect to the reference antenna. Also, $c_{x}$ and $c_{z}$ denote the spacing between two adjacent meta-atoms along the $x$- and $z$-directions, respectively, and $N_{x}$ and $N_{z}$ denote the number of meta-atoms along the $x$- and $z$- axes, respectively.

Then we can express $d_{n,m}^{1}(\hat{y}_{n}^{1})=\sqrt{\rho_{n,m}^{1}+(\bar{y}^{1}+\hat{y}_{n}^{1})^{2}},$ and $\cos(\theta_{n,m}^{1})=\frac{\bar{y}^{1}+\hat{y}_{n}^{1}}{\sqrt{\rho_{n,m}^{1}+(\bar{y}^{1}+\hat{y}_{n}^{1})^{2}}},$ where $\rho_{n,m}^{1}\triangleq(x_{n}^{1})^{2}+(z_{n}^{1}-z_{m}^{0})^{2}$. Thus 
\begin{equation}
[\boldsymbol{\Omega}^{1}(\hat{\mathbf{y}})]_{n,m}=w_{n,m}^{1}(\hat{y}_{n}^{1})q_{n,m}^{1}(\hat{y}_{n}^{1})r_{n,m}^{1}(\hat{y}_{n}^{1}),\label{eq:cm1_ref}
\end{equation}
where $w_{n,m}^{1}(\hat{y}_{n}^{1})\triangleq\frac{A_{\text{a}}(\bar{y}^{1}+\hat{y}_{n}^{1})}{\rho_{n,m}^{1}+(\bar{y}^{1}+\hat{y}_{n}^{1})^{2}}$, $q_{n,m}^{1}(\hat{y}_{n}^{1})\triangleq\frac{1}{2\pi\sqrt{\rho_{n,m}^{1}+(\bar{y}^{1}+\hat{y}_{n}^{1})^{2}}}-\frac{j}{\lambda}$ and $r_{n,m}^{1}(\hat{y}_{n}^{1})\triangleq\exp\Big(j\frac{2\pi\sqrt{\rho_{n,m}^{1}+(\bar{y}^{1}+\hat{y}_{n}^{1})^{2}}}{\lambda}\Big)$.

\paragraph{\textit{$\boldsymbol{\Omega}^{\ell} (\hat{\mathbf{y}})$ for $\ell\in\mathcal{L}\setminus\{1\}$}}

Assuming that the meta-atoms of all surfaces are perfectly aligned along the $x$-$z$ plane, and taking the element in the top-left corner of the RE of the $(\ell-1)$-th surface as a reference point, we can find the coordinates of the $m$-th meta-atom of the $(\ell-1)$-th layer as $\big[x_{m}^{\ell-1},\ y_{m}^{\ell-1},\ z_{m}^{\ell-1}\big]=\big[c_{x}+\mod (m-1,N_{x}),\ \hat{y}_{v}^{\ell-1},\ z_{1}^{1}+c_{z}\lfloor(m-1)/N_{x}\rfloor\big]$. Similarly, the coordinates of the $n$-th meta-atom of the $\ell$-th layer are $\big[x_{n}^{\ell},\ y_{n}^{\ell},\ z_{n}^{\ell}\big]=\big[c_{x}+\mod (n-1,N_{x}),\ \bar{y}^{\ell}+\hat{y}_{v}^{\ell},\ z_{1}^{1}+c_{z}\lfloor(n-1)/N_{x}\rfloor\big]$.

Hence, we can express $d_{n,m}^{\ell}(\hat{y}_{m}^{\ell-1},\hat{y}_{n}^{\ell})=\sqrt{\rho_{n,m}^{\ell}+(\bar{y}^{\ell}+\hat{y}_{n}^{\ell}-\hat{y}_{m}^{\ell-1})^{2}},$ and $\cos(\theta_{n,m}^{\ell})=\frac{\bar{y}^{\ell}+\hat{y}_{n}^{\ell}-\hat{y}_{m}^{\ell-1}}{\sqrt{\rho_{n,m}^{\ell}+(\bar{y}^{\ell}+\hat{y}_{n}^{\ell}-\hat{y}_{m}^{\ell-1})^{2}}},$ where $\rho_{n,m}^{\ell}\triangleq(x_{n}^{\ell}-x_{m}^{\ell-1})^{2}+(z_{n}^{\ell}-z_{m}^{\ell-1})^{2}$. Thus 
\begin{equation}
[\boldsymbol{\Omega}^{\ell}(\hat{\mathbf{y}})]_{n,m}\!=\!w_{n,m}^{\ell}(\hat{y}_{m}^{\ell-1},\hat{y}_{n}^{\ell})q_{n,m}^{\ell}(\hat{y}_{m}^{\ell-1},\hat{y}_{n}^{\ell})r_{n,m}^{\ell}(\hat{y}_{m}^{\ell-1},\hat{y}_{n}^{\ell}),\label{eq:cml_ref}
\end{equation}
where $w_{n,m}^{\ell}(\hat{y}_{m}^{\ell-1},\hat{y}_{n}^{\ell})\triangleq\frac{A_{\text{m}}(\bar{y}^{\ell}+\hat{y}_{n}^{\ell}-\hat{y}_{m}^{\ell-1})}{\rho_{n,m}^{\ell}+(\bar{y}^{\ell}+\hat{y}_{n}^{\ell}-\hat{y}_{m}^{\ell-1})^{2}}$, $q_{n,m}^{\ell}(\hat{y}_{m}^{\ell-1},\hat{y}_{n}^{\ell})\triangleq\frac{1}{2\pi\sqrt{\rho_{n,m}^{\ell}+(\bar{y}^{\ell}+\hat{y}_{m}^{\ell}-\hat{y}_{n}^{\ell-1})^{2}}}-\frac{j}{\lambda}$ and $r_{n,m}^{\ell}(\hat{y}_{m}^{\ell-1},\hat{y}_{n}^{\ell})\triangleq\exp\Big(j\frac{2\pi\sqrt{\rho_{n,m}^{\ell}+(\bar{y}^{\ell}+\hat{y}_{n}^{\ell}-\hat{y}_{m}^{\ell-1})^{2}}}{\lambda}\Big)$.

\paragraph{\textit{$\mathbf{h}_{k}(\hat{\mathbf{y}})$}}

We assume a multipath channel between the final layer and the $k$-th user~\cite{2025_An1}, which is expressed as 
\begin{equation}
\mathbf{h}_{k}(\hat{\mathbf{y}})=\sum\nolimits_{i=0}^{I-1}\alpha_{i,k}\mathbf{a}(\hat{\mathbf{y}},\vartheta_{i,k},\varphi_{i,k})\ \forall k\in\mathcal{K},\label{eq:SV_CM}
\end{equation}
where $\mathcal{K} \triangleq \{ 1,\dots,K \}$, $I$ represents the number of paths, with the zeroth path representing the line-of-sight (LoS) component and $i\in\{1,\dots,I-1\}$ representing the non-line-of-sight (NLoS) channels. Moreover, $\alpha_{i,k}$ is the path gain of the $i$-th path and $\mathbf{a}(\hat{\mathbf{y}},\vartheta_{i,k},\varphi_{i,k})$ is the steering vector of the $k$-th user, which is a function of the azimuth and elevation angles of departure (AoDs), $\vartheta_{i,k}$ and $\varphi_{i,k}$, of the $i$-th path from the final SFIM layer.

Taking the top-left meta-atom of the RE of the final layer as a reference element, we can express the $u$-th element of the steering vector as\footnote{The channel model in~\eqref{eq:SV_CM} assumes that the maximum morphing distance (i.e., $\tilde{y}$) is negligible compared to the communication distance between the final SFIM layer and the receivers. As a result, the path gains across all meta-atoms are approximately equal. This assumption is valid in practice, as the morphing distance is typically on the order of millimeters, whereas the communication distance is usually on the order of tens of meters.} $[\mathbf{a}(\hat{\mathbf{y}},\vartheta,\varphi)]_{u}=\exp\big(j\frac{2\pi}{\lambda}\big(\psi_{u}(\vartheta,\varphi)+\hat{y}_{u}^{L}\sin\vartheta\sin\varphi\big)\big),$ where $\psi_{u}(\vartheta,\varphi)\triangleq c_{x}\mod (u-1,N_{x})\cos\vartheta\sin\varphi+c_{z}\lfloor(u-1)/N_{x}\rfloor\cos\varphi$.

\subsection{Received Signal Model}

The signal received by the $k$-th user can expressed as $f_{k}=\mathbf{g}_{k}\trans\mathbf{x}+n_{k},$ where $n_{k}\sim\mathcal{CN}(0,\sigma_{k}^{2})$ is the additive white Gaussian noise (AWGN).

\section{Problem Formulation\label{sec:prob_form} }

We can express the SINR of the $k$-th user as 
\begin{equation}
\gamma_{k}(\mathbf{p},\hat{\mathbf{y}},\boldsymbol{\phi})=\frac{p_{k}G_{k,k}(\hat{\mathbf{y}},\boldsymbol{\phi})}{\sum\nolimits_{i\in\mathcal{K}\setminus\{k\}}p_{i}G_{k,i}(\hat{\mathbf{y}},\boldsymbol{\phi})+\sigma_{k}^{2}}.
\end{equation}
where $\mathbf{p}\triangleq \text{diag} (\mathbf{P})$ and $G_{k,i}(\hat{\mathbf{y}},\boldsymbol{\phi})\triangleq\vert\mathbf{h}_{k}\trans(\hat{\mathbf{y}})\boldsymbol{\Phi}^{L}\boldsymbol{\Omega}^{L}(\hat{\mathbf{y}})\dots\boldsymbol{\Phi}^{1}\boldsymbol{\omega}^{1}_i (\hat{\mathbf{y}})\vert^{2}$ with $\boldsymbol{\omega}^{1}_i(\hat{\mathbf{y}})$ representing the $i$-th column of $\boldsymbol{\Omega}^{1}(\hat{\mathbf{y}})$. Then, we can define the rate of the $k$-th user as 
\begin{equation}
R_{k}(\mathbf{p},\hat{\mathbf{y}},\boldsymbol{\phi})=\log_{2}\big(1+\gamma_{k}(\mathbf{p},\hat{\mathbf{y}},\boldsymbol{\phi})\big),
\end{equation}
which can be expressed as 
\begin{align}
\!\!  R_{k}(\mathbf{p},\hat{\mathbf{y}},\boldsymbol{\phi}) & =\ \log_{2} \Big( \sum\nolimits_{i \in \mathcal{K}} J_{k,i} (\mathbf{p},\hat{\mathbf{y}},\boldsymbol{\phi})\!+  \!\sigma_{k}^{2} \Big)\! \nonumber \\
\!\!  & - \Big( \sum\nolimits_{i \in \mathcal{K} \setminus \{ k\}} J_{k,i} (\mathbf{p},\hat{\mathbf{y}},\boldsymbol{\phi})\!+  \!\sigma_{k}^{2} \Big),
\end{align}
where $J_{k,i} (\mathbf{p},\hat{\mathbf{y}},\boldsymbol{\phi}) \triangleq \vert 
\mathbf{g}\trans_{k}(\hat{\mathbf{y}},\boldsymbol{\phi})\mathbf{p}_i\vert^2$ and $\mathbf{p}_i$ is a vector containing zeros everywhere except at the $i$-th entry, which is $\sqrt{p_i}$.

Hence, the sum rate is given by
\begin{equation}
R_{\text{sum}}\big(\mathbf{p},\hat{\mathbf{y}},\boldsymbol{\phi}\big)=\sum\nolimits_{k\in\mathcal{K}}R_{k}\big(\mathbf{p},\hat{\mathbf{y}},\boldsymbol{\phi}\big).
\end{equation}

Our main goal is to maximize the sum rate. This problem can be formulated as follows: 
\begin{subequations}
\label{eq:main_opt}
\begin{align}
\underset{\mathbf{p},\hat{\mathbf{y}},\boldsymbol{\Phi}}{\maximize}\  & R_{\text{sum}}\big(\mathbf{p},\hat{\mathbf{y}},\boldsymbol{\phi}\big)\label{eq:obj_fun}\\
\st\  & \sum\nolimits_{k\in\mathcal{K}}p_k \leq P_{\max},\label{eq:power_cons}\\
\  & p_{k}\geq0\ \forall k\in\mathcal{K},\label{eq:power_value_cons}\\
\  & |\phi_{n}^{\ell}\big|=1\ \forall n\in\mathcal{N},\ \forall\ell\in\mathcal{L},\label{eq:RIS_cons}\\
\  & -\tilde{y}\leq\hat{{y}}_{n}^{\ell}\leq\tilde{y}\ \forall n\in\mathcal{N},\ \forall\ell\in\mathcal{L},\label{eq:morp_cons}
\end{align}
\end{subequations}
where $P_{\text{max}}$ represents the maximum transmit power budget. In this optimization problem,~\eqref{eq:power_cons} represents the maximum transmit power constraint,~\eqref{eq:power_value_cons} ensures that the power values are not negative,~\eqref{eq:RIS_cons} imposes the unit-modulus constraint of the meta-atoms’ responses, and~\eqref{eq:morp_cons} ensures that the morphing of the meta-atoms lies within the morphing range.

Solving the optimization problem~\eqref{eq:main_opt} is challenging due to the non-convexity of the objective function and the constraint~\eqref{eq:RIS_cons}, as well as the coupling between the optimization variables in the objective function. In the following section, we provide an AO solution based on the gradient projection method to obtain a stationary solution to~\eqref{eq:main_opt}. 

\section{Proposed Solution\label{sec:prop_sol} }

In this section, we propose an AO-based solution to solve the optimization problem~\eqref{eq:main_opt} by optimizing one block of variables while keeping the others fixed. 

\subsection{Morphing Optimization}
We first assume that the values of $\mathbf{p}$ and $\boldsymbol{\phi}$ are fixed, and we aim to update $\hat{\mathbf{y}}$ employing the gradient projection method. We can obtain the gradient of $R_{k}(\hat{\mathbf{y}})$ with respect to $\hat{\mathbf{y}}$ as 
\begin{equation}
\nabla_{\hat{\mathbf{y}}}R_{k}(\hat{\mathbf{y}})=\frac{1}{\ln(2)}\Big[\tfrac{\sum\nolimits_{i \in \mathcal{K}} \nabla_{\hat{\mathbf{y}}} J_{k,i} (\hat{\mathbf{y}})\!}{\sum\nolimits_{i \in \mathcal{K}} J_{k,i} (\hat{\mathbf{y}})\!+  \!\sigma_{k}^{2}}-\tfrac{\sum\nolimits_{i \in \mathcal{K} \setminus \{k \}} \nabla_{\hat{\mathbf{y}}} J_{k,i} (\hat{\mathbf{y}})\!}{\sum\nolimits_{i \in \mathcal{K} \setminus \{k \}} J_{k,i} (\hat{\mathbf{y}})\!+  \!\sigma_{k}^{2}}\Big].
\end{equation}
Next, we find the gradient $\nabla_{\hat{\mathbf{y}}}J_{k,i}(\hat{\mathbf{y}})=\Big[\frac{\partial J_{k,i}(\hat{\mathbf{y}})}{\partial\hat{y}_{1}^{1}},\dots,\frac{\partial J_{k,i}(\hat{\mathbf{y}})}{\partial\hat{y}_{N}^{1}},\dots,\frac{\partial J_{k,i}(\hat{\mathbf{y}})}{\partial\hat{y}_{1}^{L}},\dots,\frac{\partial J_{k,i}(\hat{\mathbf{y}})}{\partial\hat{y}_{N}^{L}}\Big]\trans$.

\paragraph{For\textit{\emph{ $\ell\in\mathcal{L}\setminus\{L\}$}}}

\begin{multline}
\tfrac{\partial J_{k,i}(\hat{\mathbf{y}})}{\partial\hat{y}_{n}^{\ell}}=2\Re\Big\{\Big[\big(\mathbf{c}_{k}^{\ell}(\hat{\mathbf{y}})\big)\trans\Big\{\tfrac{\partial\boldsymbol{\Omega}^{\ell+1}(\hat{\mathbf{y}})}{\partial\hat{y}_{n}^{\ell}}\boldsymbol{\Phi}^{\ell}\boldsymbol{\Omega}^{\ell}(\hat{\mathbf{y}})\\
+\boldsymbol{\Omega}^{\ell+1}(\hat{\mathbf{y}})\boldsymbol{\Phi}^{\ell}\tfrac{\partial\boldsymbol{\Omega}^{\ell}(\hat{\mathbf{y}})}{\partial\hat{y}_{n}^{\ell}}\Big\}\mathbf{e}_i^{\ell}(\hat{\mathbf{y}})\Big]^{*}\mathbf{g}\trans_{k}(\hat{\mathbf{y}})\mathbf{p}_i\Big\},\label{eq:dGk}
\end{multline}
where $(\mathbf{c}_{k}^{\ell}(\hat{\mathbf{y}}))\trans\triangleq(\mathbf{h}_{k}(\hat{\mathbf{y}}))\trans\big(\prod_{u=L}^{\ell+2}\boldsymbol{\Phi}^{u}\boldsymbol{\Omega}^{u}(\hat{\mathbf{y}})\big)\boldsymbol{\Phi}^{\ell+1}$ and $\mathbf{e}_i^{\ell}(\hat{\mathbf{y}})\triangleq (\prod_{u=\ell-1}^{1}\boldsymbol{\Phi}^{u}\boldsymbol{\Omega}^{u}(\hat{\mathbf{y}})) \mathbf{p}_i$.

We can observe from~\eqref{eq:dGk} that only the $n$-th column of $\frac{\partial\boldsymbol{\Omega}^{\ell}(\hat{\mathbf{y}})}{\partial\hat{y}_{n}^{\ell}}$ can have nonzero elements. Therefore, the product $\boldsymbol{\Omega}^{\ell+1}(\hat{\mathbf{y}})\boldsymbol{\Phi}^{\ell}\frac{\partial\boldsymbol{\Omega}^{\ell}(\hat{\mathbf{y}})}{\partial\hat{y}_{n}^{\ell}}$ can be expressed as a matrix whose elements are all zero except for the $n$-th column, which is $\phi_{n}^{\ell}\boldsymbol{\Omega}^{\ell+1}(\hat{\mathbf{y}})\mathbf{z}_{n}^{\ell}(\hat{\mathbf{y}})$, where 
\begin{multline}
[\mathbf{z}_{n}^{\ell}(\mathbf{y})]_{m}\triangleq\tfrac{\partial w_{n,m}^{\ell}(\mathbf{y})}{\partial\hat{y}_{n}^{\ell}}q_{n,m}^{\ell}(\mathbf{y})r_{n,m}^{\ell}(\mathbf{y})+w_{n,m}^{\ell}(\mathbf{y})\\
\times\tfrac{\partial q_{n,m}^{\ell}(\mathbf{y})}{\partial\hat{y}_{n}^{\ell}}r_{n,m}^{\ell}(\mathbf{y})+w_{n,m}^{\ell}(\mathbf{y})q_{n,m}^{\ell}(\mathbf{y})\tfrac{\partial r_{n,m}^{\ell}(\mathbf{y})}{\partial\hat{y}_{n}^{\ell}},\label{eq:beg}
\end{multline}
with $\frac{\partial w_{n,m}^{\ell}(\hat{\mathbf{y}})}{\partial\hat{y}_{n}^{\ell}}=\frac{\tilde{A}^{\ell}\big(\rho_{n,m}-(\hat{y}_{n}^{\ell}+\xi_{m}^{\ell})^{2}\big)}{\big(\rho_{n,m}+(\hat{y}_{n}^{\ell}+\xi_{m}^{\ell})^{2}\big)^{2}},$ $\frac{\partial q_{n,m}^{\ell}(\hat{\mathbf{y}})}{\partial\hat{y}_{n}^{\ell}}=-\frac{\hat{y}_{n}^{\ell}+\xi_{m}^{\ell}}{2\pi\big(\rho_{n,m}+(\hat{y}_{n}^{\ell}+\xi_{m}^{\ell})^{2}\big)^{3/2}},$ and $\frac{\partial r_{n,m}^{\ell}(\hat{\mathbf{y}})}{\partial\hat{y}_{n}^{\ell}}=j\frac{2\pi(\hat{y}_{n}^{\ell}+\xi_{m}^{\ell})}{\lambda\sqrt{\rho_{n,m}+(\hat{y}_{n}^{\ell}+\xi_{m}^{\ell})^{2}}}\times\exp\Big(j\frac{2\pi\sqrt{\rho_{n,m}+(\hat{y}_{n}^{\ell}+\xi_{m}^{\ell})^{2}}}{\lambda}\Big).$ Moreover, 
\begin{equation}
\xi_{m}^{\ell}\triangleq\begin{cases}
\bar{y}^{1}, & \ell=1,\\
\bar{y}^{\ell}-\hat{y}_{m}^{\ell-1}, & \ell\in \mathcal{L} \setminus \{1\}.
\end{cases}\label{eq:end}
\end{equation}

Similarly, only the $n$-th row of $\frac{\partial\boldsymbol{\Omega}^{\ell+1}(\hat{\mathbf{y}})}{\partial\hat{y}_{n}^{\ell}}$ can have nonzero elements. Therefore, the product $\frac{\partial\boldsymbol{\Omega}^{\ell+1}(\hat{\mathbf{y}})}{\partial\hat{y}_{n}^{\ell}}\boldsymbol{\Phi}^{\ell}\boldsymbol{\Omega}^{\ell}(\hat{\mathbf{y}})$ can be expressed as a matrix whose elements are all zero except for the $n$-th row, which is $\phi_{n}^{\ell}\big(\boldsymbol{\eta}_{n}^{\ell}(\hat{\mathbf{y}})\big)\trans\boldsymbol{\Omega}^{\ell}(\hat{\mathbf{y}})$, where 
\begin{multline}
[\boldsymbol{\eta}_{n}^{\ell}(\hat{\mathbf{y}})]_{m}\triangleq\tfrac{\partial w_{m,n}^{\ell+1}(\hat{\mathbf{y}})}{\partial\hat{y}_{n}^{\ell}}q_{m,n}^{\ell+1}(\hat{\mathbf{y}})r_{m,n}^{\ell+1}(\hat{\mathbf{y}})+w_{m,n}^{\ell+1}(\hat{\mathbf{y}})\\
\times\tfrac{\partial q_{m,n}^{\ell+1}(\hat{y}_{n}^{\ell})}{\partial\hat{y}_{n}^{\ell}}r_{m,n}^{\ell+1}(\hat{y}_{n}^{\ell})+w_{m,n}^{\ell+1}(\hat{y}_{n}^{\ell})q_{m,n}^{\ell+1}(\hat{y}_{n}^{\ell})\tfrac{\partial r_{m,n}^{\ell+1}(\hat{y}_{n}^{\ell})}{\partial\hat{y}_{n}^{\ell}},
\end{multline}
with $\frac{\partial p_{m,n}^{\ell+1}(\hat{y}_{n}^{\ell})}{\partial\hat{y}_{n}^{\ell}}=\frac{\tilde{A}^{\ell}\big((\hat{y}_{n}^{\ell}-\varsigma_{m}^{\ell})^{2}-\rho_{m,n}\big)}{\big(\rho_{m,n}+(\hat{y}_{n}^{\ell}-\varsigma_{m}^{\ell})^{2}\big)^{2}},$ $\frac{\partial q_{m,n}^{\ell+1}(\hat{\mathbf{y}})}{\partial\hat{y}_{n}^{\ell}}=-\frac{\hat{y}_{n}^{\ell}-\varsigma_{m}^{\ell}}{2\pi\big(\rho_{m,n}+(\hat{y}_{n}^{\ell}-\varsigma_{m}^{\ell})^{2}\big)^{3/2}},$ $\frac{\partial r_{m,n}^{\ell+1}(\hat{\mathbf{y}})}{\partial\hat{y}_{n}^{\ell}}=j\frac{2\pi(\hat{y}_{n}^{\ell}-\varsigma_{m}^{\ell})}{\lambda\sqrt{\rho_{m,n}+(\hat{y}_{n}^{\ell}-\varsigma_{m}^{\ell})^{2}}}\times\exp\big(j2\pi\sqrt{\rho_{m,n}+(\hat{y}_{n}^{\ell}-\varsigma_{m}^{\ell})^{2}}\big/\lambda\big),$ and $\varsigma_{m}^{\ell}\triangleq\bar{y}^{\ell+1}+\hat{y}_{m}^{\ell+1}.$

\paragraph{For $\ell=L$}

\begin{multline}
\tfrac{\partial J_{k,i}(\hat{\mathbf{y}})}{\partial\hat{y}_{n}^{\ell}}=2\Re\Big\{\Big[\Big\{\Big(\tfrac{\partial\mathbf{h}_{k}(\hat{\mathbf{y}})}{\partial\hat{y}_{n}^{L}}\Big)\trans\boldsymbol{\Phi}^{L}\boldsymbol{\Omega}^{L}(\hat{\mathbf{y}})\\
+\mathbf{h}_{k}\trans(\hat{\mathbf{y}})\boldsymbol{\Phi}^{L}\tfrac{\partial\boldsymbol{\Omega}^{L}(\hat{\mathbf{y}})}{\partial\hat{y}_{n}^{L}}\Big\}\mathbf{e}_i^{\ell}(\hat{\mathbf{y}})\Big]^{*}\mathbf{g}\trans_{k}(\hat{\mathbf{y}})\mathbf{p}_i\Big\},
\end{multline}

On the one hand, we can obtain $\frac{\partial\boldsymbol{\Omega}^{L}(\hat{\mathbf{y}})}{\partial\hat{y}_{n}^{L}}$ using~\eqref{eq:beg} and~\eqref{eq:end}. On the other hand, we can obtain $\frac{\partial\mathbf{h}_{k}(\hat{\mathbf{y}})}{\partial\hat{y}_{n}^{L}}$ as 

\begin{multline}
\tfrac{\partial\mathbf{h}_{k}(\hat{\mathbf{y}})}{\partial\hat{y}_{n}^{\ell}}=j\tfrac{2\pi}{\lambda}\sum\nolimits_{i=0}^{I-1}\alpha_{i,k}\sin\vartheta_{i,k}\sin\varphi_{i,k}\exp\big(j\tfrac{2\pi}{\lambda}\\
\times\big\{\psi_{n}(\vartheta_{i,k},\varphi_{i,k})+\hat{y}_{n}^{L}\sin\vartheta_{i,k}\sin\varphi_{i,k}\big\}\big),
\end{multline}

Then, we can find $\nabla_{\hat{\mathbf{y}}}R_{\text{sum}}(\hat{\mathbf{y}})$ as 
\begin{equation}
\nabla_{\hat{\mathbf{y}}}R_{\text{sum}}\big(\hat{\mathbf{y}}\big)=\sum\nolimits_{k\in\mathcal{K}}\nabla_{\hat{\mathbf{y}}}R_{k}\big(\hat{\mathbf{y}}\big).
\end{equation}
Thus, we can update $\hat{\mathbf{y}}$ as follows: 
\begin{equation}
\hat{\mathbf{y}}^{(t)}\leftarrow\hat{\mathbf{y}}^{(t-1)}+\delta_{\hat{\mathbf{y}}}\nabla_{\hat{\mathbf{y}}}R_{\text{sum}}(\hat{\mathbf{y}}^{(t-1)}),\label{eq:y_update}
\end{equation}
where $\delta_{\hat{\mathbf{y}}}$ is the step size and $\hat{\mathbf{y}}^{(t)}$ is the value of $\hat{\mathbf{y}}$ at the $t$-th iteration. Then the projection onto the feasible set can be performed as follows: 
\begin{equation}
\hat{\mathbf{y}}^{(t)}\leftarrow\max\big\{-\tilde{y}\mathbf{1},\min\big\{\hat{\mathbf{y}}^{(t)},\tilde{y}\mathbf{1}\big\}\big\}.\label{eq:proj_y}
\end{equation}

\subsection{Power Allocation}
Next, we assume that the values of $\hat{\mathbf{y}}$ and $\boldsymbol{\phi}$ are fixed. We can re-express the sum rate as
\begin{align}
\!\!  R_{k}(\mathbf{p}) & =\ \sum\nolimits_{k \in \mathcal{K}} \log_{2}(\Vert 
\mathbf{G}_{k}\mathbf{p}\Vert_2^2\!+  \!\sigma_{k}^{2})\! \nonumber \\
\!\!  & -\!\log_{2}(\Vert \mathbf{G}_{k}\tilde{\mathbf{I}}_k\mathbf{p}\Vert_2^2\!+\!\sigma_{k}^{2}),
\label{eq:Rk}
\end{align}
where $\tilde{\mathbf{I}}_{k}$ is an identity matrix with the $(k,k)$-th entry set to zero, while all other diagonal entries remain one and $\mathbf{G}_{k} \triangleq \text{diag} (\mathbf{g}_{k})$. Therefore, We can find the gradient $\nabla_{\mathbf{p}}R_{\text{sum}}(\mathbf{p})$ as 
\begin{equation}
\nabla_{\mathbf{p}}R_{\text{sum}}(\mathbf{p})=\frac{2}{\ln(2)}\sum\nolimits_{k\in\mathcal{K}}\Big(\tfrac{\mathbf{G}^*_k\mathbf{G}_{k}\mathbf{p}}{\Vert 
\mathbf{G}_{k}\mathbf{p}\Vert_2^2\!+  \!\sigma_{k}^{2}}-\tfrac{\tilde{\mathbf{I}}_k \mathbf{G}^*_{k} \mathbf{G}_{k} \tilde{\mathbf{I}}_k \mathbf{p}}{\Vert 
\mathbf{G}_{k} \tilde{\mathbf{I}}_k \mathbf{p}\Vert_2^2\!+  \!\sigma_{k}^{2}}\Big).
\end{equation}

Then, we can update the power allocation as 
\begin{equation}
\mathbf{p}^{(t)}\leftarrow\mathbf{p}^{(t-1)}+\delta_{\mathbf{p}}\nabla_{\mathbf{p}}R_{\text{sum}}\big(\mathbf{p}^{(t-1)}\big),\label{eq:update_P}
\end{equation}
where $\delta_{\mathbf{p}}$ is the step size. Then, we can project onto the feasible set using the following projection rules sequentially: 
\begin{equation}
\mathbf{p}^{(t)}\leftarrow\max\Big\{\mathbf{0},\mathbf{p}^{(t)}\Big\},\label{eq:proj_P1}
\end{equation}
followed by 
\begin{equation}
\mathbf{p}^{(t)}\leftarrow\min\Big\{\mathbf{p}^{(t)},\tfrac{\sqrt{P_{\text{max}}}}{\Vert\mathbf{p}^{(t)} \Vert_2}\mathbf{1}\Big\}.\label{eq:proj_P2}
\end{equation}

\begin{algorithm}[t]
\caption{Sum rate maximization for SFIM-based system.}
\label{algo1}

\KwIn{ $\mathbf{p}^{(0)}$, $\boldsymbol{\Phi}^{(0)}$, $\hat{\mathbf{y}}^{(0)}$, $\epsilon\geq0$, $t=1$ }

\Repeat{ $\big|R_{\mathrm{sum}}(\mathbf{p}^{(t)},\boldsymbol{\Phi}^{(t)},\hat{\mathbf{y}}^{(t)})-R_{\mathrm{sum}}(\mathbf{p}^{(t-1)},\boldsymbol{\Phi}^{(t-1)},\hat{\mathbf{y}}^{(t-1)})\big|\leq\epsilon$}{

Update $\hat{\mathbf{y}}^{(t)}$ using~\eqref{eq:y_update} and project onto the feasible set using~\eqref{eq:proj_y}\;

Update $\mathbf{p}^{(t)}$ using~\eqref{eq:update_P} and project onto the feasible set using~\eqref{eq:proj_P1} then~\eqref{eq:proj_P2}\;

Update ${\boldsymbol{\Phi}}^{(t)}$ using~\eqref{eq:Phi_update} and project onto the feasible set using~\eqref{eq:proj_Phi}\;

$t\leftarrow t+1$\;

}

\KwOut{ $\mathbf{p}=\mathbf{p}^{(t)}$, $\boldsymbol{\Phi}=\boldsymbol{\Phi}^{{(t)}}$, $\hat{\mathbf{y}}=\hat{\mathbf{y}}^{(t)}$ }
\end{algorithm}

\subsection{Meta-Atom Response Optimization}

Next, we assume that the values of $\hat{\mathbf{y}}$ and $\mathbf{p}$ are fixed. Then, we find $\nabla_{\boldsymbol{\phi}}R_{\text{sum}}(\boldsymbol{\phi})$ as 
\begin{align}
& \nabla_{\boldsymbol{\phi}}R_{\text{sum}}(\boldsymbol{\phi}) \notag \\
= & \  \frac{1}{\ln(2)}\sum_{k \in \mathcal{K}} \Big[\tfrac{\sum\nolimits_{i \in \mathcal{K}} \nabla_{\boldsymbol{\phi}} J_{k,i} (\boldsymbol{\phi})\!}{\sum\nolimits_{i \in \mathcal{K}} J_{k,i} (\boldsymbol{\phi})\!+  \!\sigma_{k}^{2}} -\tfrac{\sum\nolimits_{i \in \mathcal{K} \setminus \{k \}} \nabla_{\boldsymbol{\phi}} J_{k,i} (\boldsymbol{\phi})\!}{\sum\nolimits_{i \in \mathcal{K} \setminus \{k \}} J_{k,i} (\boldsymbol{\phi})\!+  \!\sigma_{k}^{2}}\Big].
\end{align}
Next, we find $\nabla_{\boldsymbol{\phi}}J_{k,i} (\boldsymbol{\phi})=[\nabla_{\boldsymbol{\phi}^{1}}J_{k,i}\trans (\boldsymbol{\phi}),\dots,\nabla_{\boldsymbol{\phi}^{L}}J_{k,i}\trans (\boldsymbol{\phi})]\trans$, where $\boldsymbol{\phi}^{\ell}\triangleq\text{diag}(\boldsymbol{\Phi}^{L})$, as follows: 
\begin{align}
& \nabla_{\boldsymbol{\phi}^{\ell}}J_{k,i}\trans (\boldsymbol{\phi}) =\nabla_{\boldsymbol{\phi}^{\ell}}\vert(\boldsymbol{\phi}^{\ell})\trans\boldsymbol{\Xi}^{\ell}(\boldsymbol{\phi})\boldsymbol{\zeta}_i^{\ell}(\boldsymbol{\phi})\vert^{2} \notag \\
=\  & 2\boldsymbol{\Xi}^{\ell}(\boldsymbol{\phi})\boldsymbol{\zeta}_i^{\ell}(\boldsymbol{\phi})(\boldsymbol{\zeta}_i^{\ell}(\boldsymbol{\phi}))\herm(\boldsymbol{\Xi}^{\ell}(\boldsymbol{\phi}))\herm(\boldsymbol{\phi}^{\ell})^{*},
\end{align}
where $\boldsymbol{\Xi}^{\ell}(\boldsymbol{\phi})\triangleq\text{diag}(\mathbf{h}_{k}\trans\prod_{i=L}^{\ell+1}\boldsymbol{\Phi}^{i}\boldsymbol{\Omega}^{i})$ and $\boldsymbol{\zeta}_i^{\ell}(\boldsymbol{\phi})\triangleq\boldsymbol{\Omega}^{\ell}(\prod_{i=\ell-1}^{1}\boldsymbol{\Phi}^{i}\boldsymbol{\Omega}^{i})\mathbf{p}_i$.

We then use the following update rule, which leads to improved convergence behavior as discussed in~\cite{2025_Bahingayi}: 
\begin{equation}
\boldsymbol{\phi}^{(t)}\leftarrow\boldsymbol{\phi}^{(t-1)}+\delta_{\boldsymbol{\phi}}\tfrac{\nabla_{\boldsymbol{\phi}}R_{\text{sum}}(\boldsymbol{\phi}^{(t-1)})}{\big|\big[\text{diag}\big(\nabla_{\boldsymbol{\phi}}R_{\text{sum}}\big(\boldsymbol{\phi}^{(t-1)}\big)\big)\big]^{-1}\big|},\label{eq:Phi_update}
\end{equation}
where $\delta_{\boldsymbol{\phi}}$ is the step size. Then, the projection onto the feasible set can be performed as follows: 
\begin{equation}
\boldsymbol{\phi}^{(t)}\leftarrow\tfrac{\boldsymbol{\phi}^{(t)}}{\big|\big(\text{diag}\big(\boldsymbol{\phi}^{(t)}\big)\big)^{-1}\big|}.\label{eq:proj_Phi}
\end{equation}
The proposed method in summarized in \textbf{Algorithm~\ref{algo1}}.

\section{Numerical Results and Discussion\label{sec:sim} }

In this section, we present numerical simulation results to demonstrate the performance of the proposed system and compare it with that of the RSIM and hybrid SIM (HSIM) systems, where the latter employs a rigid SIM except for the final layer which is flexible.

\begin{figure}[tb]
\begin{centering}
\includegraphics[width=0.75\columnwidth]{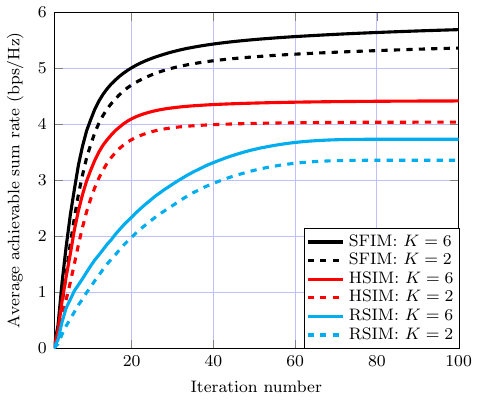}
\par\end{centering}
\caption{Average achievable sum rate at each iteration for $K=2,6$.}
\label{fig:conv}
\end{figure}

\begin{figure}[tb]
\begin{centering}
\includegraphics[width=0.75\columnwidth]{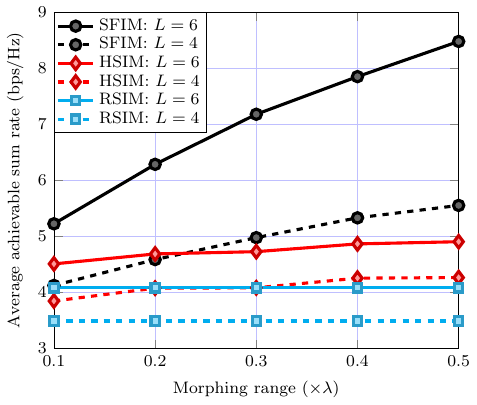}
\par\end{centering}
\caption{Average achievable sum rate as the morphing range varies for $L=4,6$.}
\label{fig:morph}
\end{figure}

Unless stated otherwise, the following simulation parameters are used. The carrier frequency is set to 28 GHz (corresponding to operation in the 3GPP NR FR2 band), which results in a carrier wavelength of $\lambda \approx \unit[10.7]{mm}$. The number of SIM layers is $L=4$, and each layer consists of $N=8\times8$ meta-atoms. The number of users is $K=4$. The spacing between the transmit antenna array and the RE of the first layer, as well as the inter-layer spacing between the REs, is $\bar{y}^{\ell}=\unit[10\lambda/L]{m}$ for each $\ell\in \mathcal{L}$. The area of a transmit antenna and that of a meta-atom are both assumed to be $A_{\text{a}}=A_{\text{m}}=\unit[\lambda^{2}/4]{m^2}$. The transmit power budget is $\unit[25]{dBm}$ and the maximum morphing distance is $\tilde{y}=\unit[\lambda/2]{m}$. Moreover, the noise power for each user $k \in \mathcal{K}$ is set to $\sigma_{k}^{2}=\unit[-104]{dBm}$. The number of paths in~\eqref{eq:SV_CM} is set to $I=6$ (i.e., one LoS and five NLoS components) and a path-loss exponent of $3.5$ is used. The distance between the reference element in the final layer and the $k$-th user is assumed to be uniformly distributed between $\unit[95]{m}$ and $\unit[105]{m}$. The azimuth and elevation AoDs for the $k$-th user are assumed to be uniformly distributed between $\unit[-\pi/4]{rad}$ and $\unit[\pi/4]{rad}$. The five scatterers are assumed to have distances and angles that are uniformly distributed between $\unit[50]{m}$ and $\unit[105]{m}$, and between $\unit[-\pi/2]{rad}$ and $\unit[-\pi/4]{rad}$, respectively.

\begin{figure}[tb]
\begin{centering}
\includegraphics[width=0.75\columnwidth]{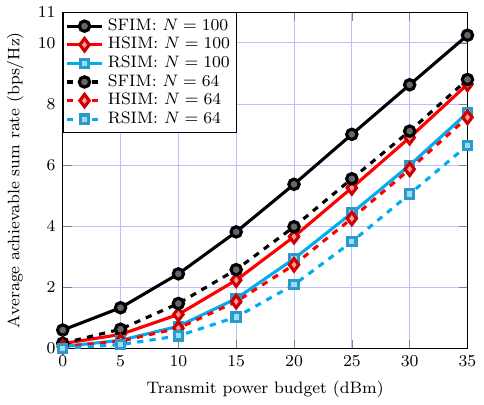}
\par\end{centering}
\caption{Average achievable sum rate as the transmit power budget varies for $N=64,100$.}
\label{fig:P_max}
\end{figure}

\begin{figure*}

\subfloat[Layer 1.]{\includegraphics[width=0.25\textwidth]{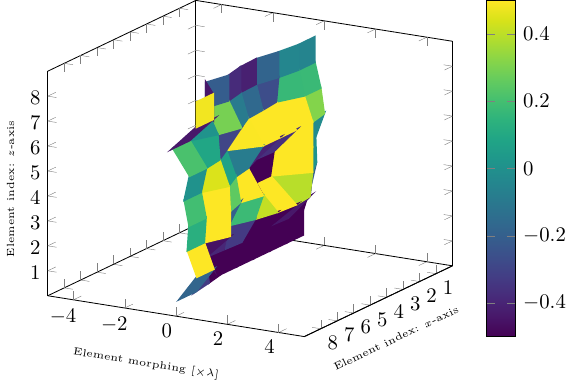}}%
\subfloat[Layer 2.]{\includegraphics[width=0.25\textwidth]{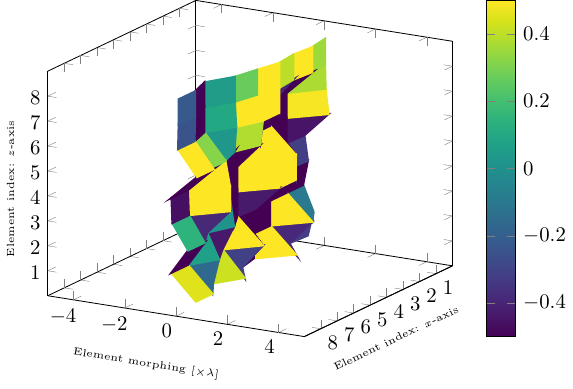}}%
\subfloat[Layer 3.]{\includegraphics[width=0.25\textwidth]{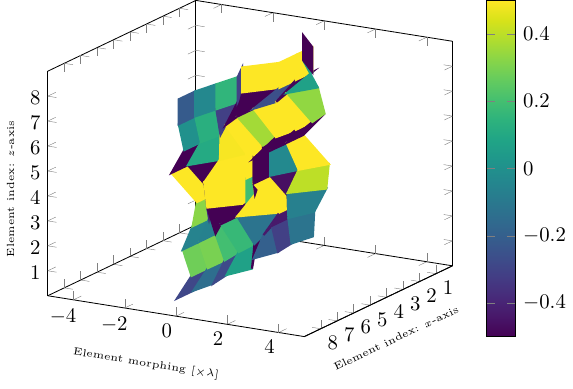}}%
\subfloat[Layer 4.]{\includegraphics[width=0.25\textwidth]{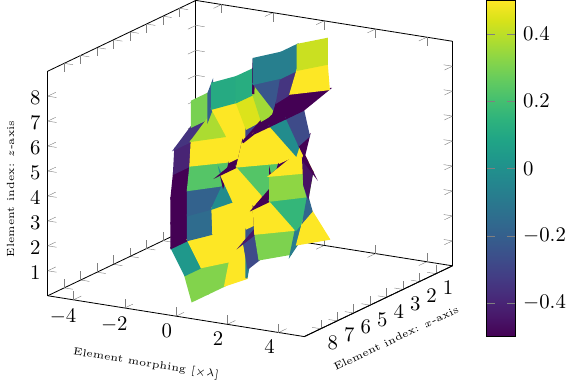}}

\caption{Example morphing of each layer after applying the proposed algorithm.}\label{fig:HM}

\end{figure*}


Fig.~\ref{fig:conv} shows the average achievable sum rate at each iteration for $K=2$ and $K=6$. We can notice that the proposed SFIM-based system offers a 47\%–58\% and a 24\%–27\% improvement in the sum rate compared to the RSIM- and HSIM-based systems, respectively. We can also observe an increase in the sum rate when the number of users is $6$ as compared to $4$. This is because the system benefits from the additional multiuser diversity, allowing more efficient resource allocation and interference management across users.

Fig.~\ref{fig:morph} depicts the average achievable sum rate as the morphing range varies for $L=4$ and $L=6$. Since the RSIM does not utilize any flexibility, its performance remains constant. We can observe an improvement in performance when the final layer is morphed (i.e., the HSIM-based system). This improvement is more pronounced for the proposed SFIM system, as all layers benefit from the additional degrees of freedom provided by a larger morphing range. The improvement varies from \unit[0.5]{bps/Hz} to \unit[2.1]{bps/Hz} for a four-layer SIM and from \unit[1.1]{bps/Hz} to \unit[4.3]{bps/Hz} for a six-layer SIM. The results indicate that the achievable sum rate can be nearly doubled for a six-layer SIM when an SFIM with a maximum morphing distance of $0.5\lambda$ is employed compared to its RSIM counterpart.

Fig.~\ref{fig:P_max} shows the average achievable sum rate as the transmit power budget increases for $N=64$ and $N=100$. We can notice that, for all transmit power budget values, the SFIM with 64 meta-atoms outperforms the 100-meta-atom HSIM and RSIM systems, thanks to the additional flexibility of the SFIM. This indicates that fewer hardware resources can be utilized with the SFIM without compromising performance. Furthermore, employing the SFIM can lead to significant power savings. For instance, for a target sum rate of \unit[7]{bps/Hz} and $N=100$, approximately \unit[9]{dBm} and \unit[5]{dBm} of power can be saved compared to the RSIM and HSIM systems, respectively.

Finally, Fig.~\ref{fig:HM} depicts an example of the optimized morphing distance for each meta-atom in the SFIM-based system. It can be observed that the morphing pattern varies across layers, highlighting the adaptability of the proposed structure. This flexibility enables the system to efficiently adapt to varying user locations and channel conditions, ultimately enhancing overall performance and robustness.

\section{Conclusion\label{sec:conc} }

This work is the first work to investigate the design of an SFIM for a multi-user communication system, where the metasurface layers are deformable to improve theoverall sum rate performance. The system model was first presented, including the channel model and the signal model, and an optimization problem was formulated to maximize the system sum rate under constraints on the transmit power budget, morphing distance, and the unit-modulus constraint of the meta-atom responses. An alternating optimization framework based on the gradient projection method was developed to solve the problem efficiently. Simulation results showed that the proposed SFIM-based system can achieve more than 50\% improvement in the sum rate compared to its rigid SIM counterpart.

\bibliographystyle{IEEEtran}
\bibliography{Bibliography}

\end{document}